\documentclass{article}
\usepackage{spconf,amsmath,graphicx}
\usepackage{amssymb}
\usepackage{bbm}
\usepackage{enumitem}
\usepackage{booktabs}

\usepackage[utf8]{inputenc} 
\usepackage[T1]{fontenc}    
\usepackage{hyperref}       
\usepackage{url}            
\usepackage{booktabs}       
\usepackage{amsfonts}       
\usepackage{nicefrac}       
\usepackage{microtype}      
\usepackage[table,xcdraw]{xcolor}   
\usepackage{graphicx}
\usepackage{color}
\usepackage{amsmath}
\usepackage{algorithm}
\usepackage{algorithmic}
\usepackage{needspace}
\usepackage{enumitem}
\usepackage{wrapfig}
\usepackage{caption}
\usepackage{multicol}
\usepackage{multirow}

\definecolor{Gray}{gray}{0.9}
\definecolor{brightturquoise}{rgb}{0.85, 1, 1}

\title{Stutter-Solver: End-to-end Multi-lingual Dysfluency Detection}
\name{\parbox{\linewidth}{\centering Xuanru Zhou$^1$, Cheol Jun Cho$^2$, Ayati Sharma$^2$, Brittany Morin$^3$, David Baquirin$^3$, Jet Vonk$^3$, Zoe Ezzes$^3$, Zachary Miller$^3$, Boon Lead Tee$^3$, Maria Luisa Gorno-Tempini$^3$, Jiachen Lian$^{2\dagger}$, Gopala Anumanchipalli$^{2\dagger}$}}

\address{
   $^1$ Zhejiang University 
   \quad $^2$ UC Berkeley 
   \quad $^3$ UCSF
   \\
   \small \tt xuanruzhou15@gmail.com,  \{jiachenlian, gopala\}@berkeley.edu
}

\begin{document}
\maketitle
\begin{abstract}
Current de-facto dysfluency modeling methods~\cite{lian2023unconstrained-udm, lian2024hierarchical} utilize template matching algorithms which are not generalizable to out-of-domain real-world dysfluencies across languages, and are not scalable with increasing amounts of training data. To handle these problems, we propose \textit{Stutter-Solver}: an end-to-end framework that detects dysfluency with accurate type and time transcription, inspired by the YOLO~\cite{redmon2016look} object detection algorithm. \textit{Stutter-Solver} can handle \textit{co-dysfluencies} and is a natural multi-lingual dysfluency detector. To leverage scalability and boost performance, we also introduce three novel dysfluency corpora: \textit{VCTK-Pro}, \textit{VCTK-Art}, and \textit{AISHELL3-Pro}, simulating natural spoken dysfluencies including repetition, block, missing, replacement, and prolongation through articulatory-encodec and TTS-based methods. Our approach achieves \textit{state-of-the-art} performance on all available dysfluency corpora.
Code and datasets are open-sourced at \url{https://github.com/eureka235/Stutter-Solver}.

\end{abstract}


\begin{figure*}[htb]
\begin{minipage}[b]{0.78\linewidth}
  \centering
 \centerline{\includegraphics[width=13.8cm]{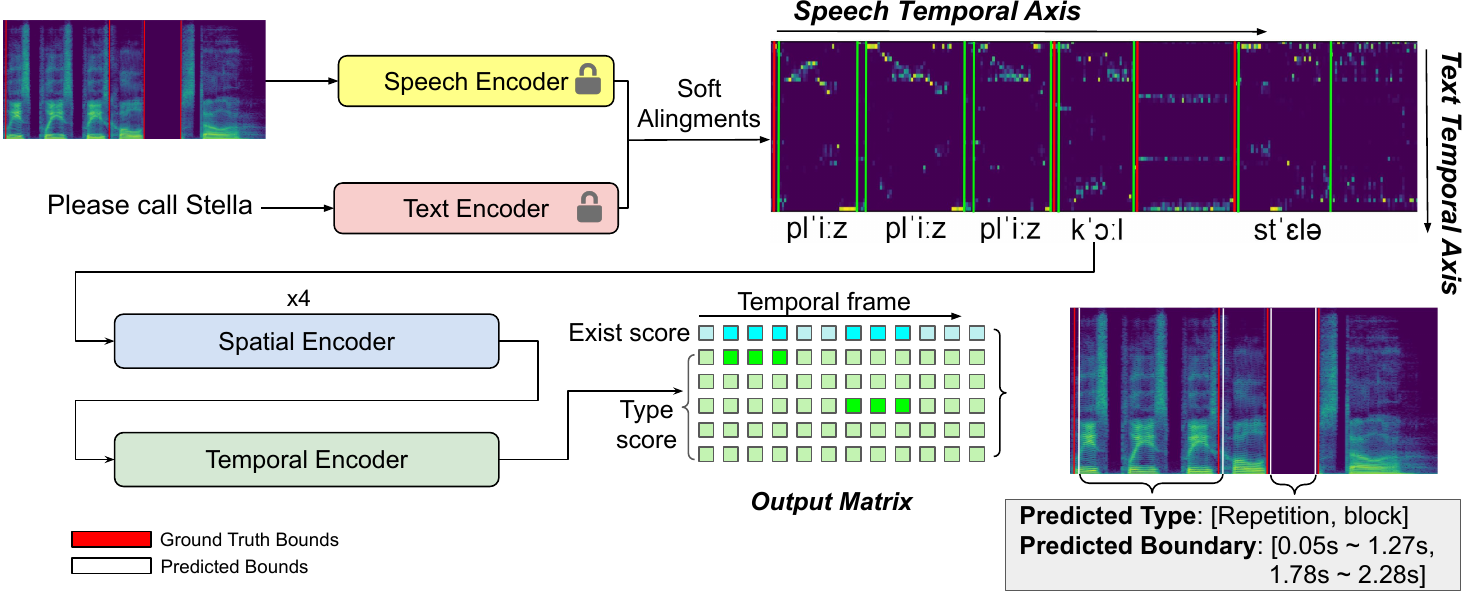}}
  \centerline{(a) Stutter-Solver architecture}\medskip
\end{minipage}
\hfill
\begin{minipage}[b]{0.2\linewidth}
  \centering
  \centerline{\includegraphics[width=2.5cm]{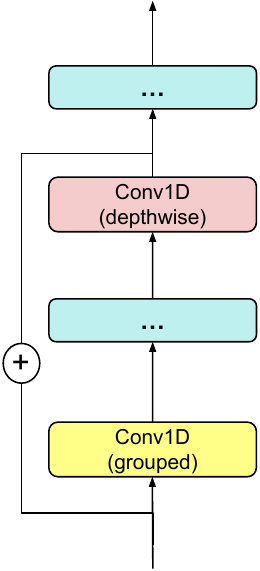}}
  \centerline{(b) Spatial encoder}\medskip
\end{minipage}
\caption{We utilize the pretrained VITS speech and text encoders to process spectrogram and reference text respectively, generating the soft speech-text alignments passed into the detector.
The output matrix contains exist confidence score and 5 types of type confidence scores (start \& end bounds are left out in the paradigm). The higher the brightness, the higher the score, indicating the existence and type of dysfluency.
\textbf{a)} shows the series nature of our detector with spatial encoder and subsequent temporal encoder. \textbf{b)} is a diagram for a spatial encoder block - grouped convolutions are important for extracting local spatial features without completely collapsing information across the text dimension.}
\label{architecture}
\end{figure*}
\section{Introduction}
\label{sec:intro}
Speech dysfluency modeling is the core module in speech therapy. The U.S. speech therapy market is projected to reach USD 6.93 billion by 2030~\cite{ppa-market}. Technically, speech dysfluency modeling is a speech recognition problem, which is dominated by large-scale developments~\cite{radford2022whisper, zhang2023google-usm, pratap2023scaling-speech, lian2023av-data2vec}. However, those large ASR models struggle with dysfluent speech because ASR is a dysfluency removal process. For a long time, researchers have mainly treated it as a classification problem. Early methods relied on hand-crafted features~\cite{CHIAAI20122157, LPCC, ESMAILI2016104, 10068490, Detect-lstm}. More recently, end-to-end classification tasks have been developed~\cite{kourkounakis2021fluentnet, alharbi2020segment-detection3, segment-detection4, howell1995automatic, 10094692}. However, two big problems remain. First, dysfluency depends on the text, which previous methods have ignored. Second, simple classification is too basic to be deployed in real speech therapy systems. ~\cite{lian2023unconstrained-udm} proposed 2D-alignment, the alignment between reference text and phoneme-level forced alignment. Then, the template matching algorithm (with each dysfluency type as a template) was performed for dysfluency detection. The subsequent work, H-UDM ~\cite{lian2024hierarchical}, proposed recursive UDM ~\cite{lian2023unconstrained-udm} that updates word boundary segments together with alignment prediction. 
Nevertheless, these methods still exhibit certain limitations. Firstly, UDM~\cite{lian2023unconstrained-udm} and H-UDM~\cite{lian2024hierarchical} are essentially feature engineering approaches, which may not adequately handle real-world dysfluencies that do not conform to predefined templates. Secondly, developing templates for each language is impractical, as dysfluency templates are inherently language-dependent. Thirdly, template matching algorithms do not utilize training data, rendering them non-scalable with respect to dysfluency data whenever it is available.

To handle the aforementioned limitations, we approach dysfluency modeling from a \textit{simple and new} perspective. Dysfluency modeling can be regarded as an \textit{object detection} problem in the 1D domain. As such, we conceptualize this as a detection task, inspired by YOLO~\cite{redmon2016look}. We propose \textit{Stutter-Solver}, which takes dysfluent speech and reference ground truth text as input, and directly predicts dysfluency types and time boundaries in an \textit{end-to-end} manner. 
Note that ~\cite{zhou2024yolo} uses the similar idea. However, \textit{Stutter-Solver} focuses on: \textit{co-dysfluencies, multi-linguality, articulatory-simulation and co-dysfluency TTS-simulation}. 
\textit{Stutter-Solver} requires high-quality annotated dysfluency data (with precisely annotated type and time boundaries).Therefore, we propose an innovative dysfluency simulation method: articulatory-based \cite{articulatory_encodec}, and we performed comparative experiments with TTS-based methods. We developed three synthetic dysfluency datasets: \textit{VCTK-Pro} and \textit{AISHELL3-Pro}, using VITS~\cite{kim2021conditional-vits} for the TTS-based method; additionally, \textit{VCTK-Art} using Articulatory Encodec ~\cite{articulatory_encodec} as a vocal tract articulation simulation tool. Both \textit{VCTK-Pro} and \textit{VCTK-Art} build upon the VCTK corpus~\cite{yamagishi2019cstr-vctk}, whereas \textit{AISHELL3-Pro} builds upon the AISHELL3 corpus~\cite{AISHELL-3_2020}. These datasets include repetition, missing, block, replacement, and prolongation at phoneme \& word levels for English, and at the character-level for Mandarin. As such, \textit{Stutter-Solver} is naturally a \textit{multi-lingual} \textit{co-dysfluency} detector with no hand-crafted templates involved. As part of the speech therapy process, we have 38 English and 8 Chinese Mandarin-speaking nfvPPA subjects~\cite{gorno2011classification-nfvppa} from clinical collaborations. The proposed \textit{Stutter-Solver} achieved state-of-the-art accuracy, bound loss, and time F1 score on our new benchmark (VCTK-Art, VCTK-Pro, AISHELL3-Pro), public corpus, and nfvPPA speech. 


\vspace{10pt}
\section{Articulatory-based Simulation}
Previous research on dysfluent speech simulation~\cite{lian2023unconstrained-udm, kourkounakis2021fluentnet} has focused on direct manipulation of waveform, which has resulted in poor naturalness, evidenced in Table.~\ref{dysfluency-stats-vctk-tts}. To address this limitation, we perform simulation in two orthogonal spaces: articulatory space and textual space. This section details \textit{articulatory-based simulation}, while section~\ref{sec:tts-simulation} elaborates on the textual space approach (\textit{TTS-based simulation}).

For \textit{articulatory-based} method, we simulate dysfluency by directly editing the articulatory control space, by utilizing an offline articulatory inversion and synthesis models (Articulatory Encodec \cite{articulatory_encodec}). The Articulatory Encodec is composed of acoustic-to-articulatory inversion (AAI) model and an articulatory vocoder. 
\cite{articulatory_encodec} shows that the articulatory encodec can successfully applied to arbitrary accents and speaker identities with high-performance. The pipeline is detailed below. 

\subsection{Method Pipeline}
\begin{figure*}[ht]
    \centering
    \includegraphics[height=4.3cm]{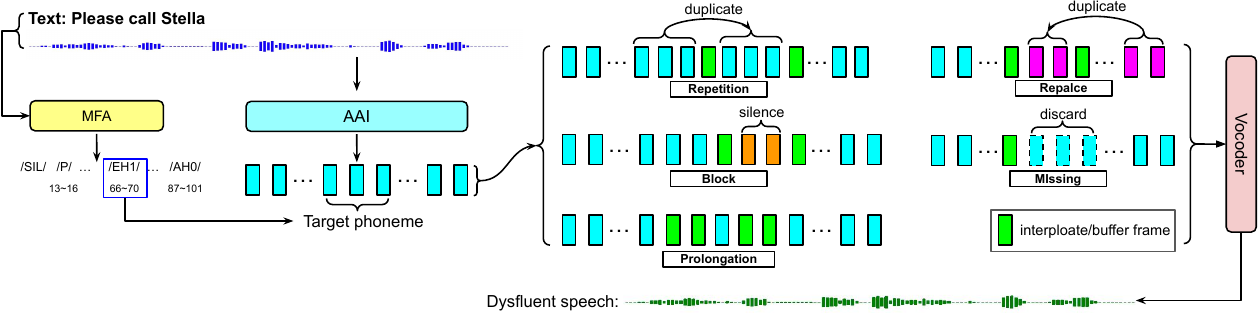}
    \caption{Pipeline of articulatory-based simulation.}
    \label{art-simulation}
\end{figure*}

We first run MFA to align raw VCTK speech with its ground truth text, obtaining 50 Hz phoneme-level force alignment that matches the EMA features from the AAI module. Various types of dysfluency are then introduced by editing the EMA features:
\textbf{Repetition:} The target phoneme segment is duplicated 2-4 times.
\textbf{Replace:} We sample a random phoneme from the current EMA feature to replace the target phoneme.
\textbf{Block:} A silence frame with 10-15 units is inserted after the target phoneme, with the silence frames sampled from the beginning of the current EMA feature.
\textbf{Missing:} The target phoneme is removed.
\textbf{Prolongation:} Interpolating within the target phoneme, extending its duration by 4 to 6 times its original length. For repetition and prolongation, the target phonemes are respectively the first phoneme of a randomly selected word and a randomly chosen vowel. For other dysfluency types, the target phonemes are selected arbitrarily without any other restrictions. To ensure smooth auditory perception, we insert a 2-unit interpolate buffer frame before and after each modification. All the interpolation operations mentioned above use bilinear interpolation. Besides phoneme-level modifications above, we also implemented \textbf{word-level repetition} and \textbf{missing}, where the target word is modified instead of the target phoneme, with all other aspects remaining identical.
The whole pipeline is depicted in Fig. \ref{art-simulation}.

Note that only the English version of articulatory-encodec model is available at this time, limiting our simulation contribution to English. However, we explored multi-lingual simulation in \textit{TTS-based Simulation}, detailed in the next section.

\section{Multi-Lingual TTS-based Simulation}
\label{sec:tts-simulation}
\subsection{Method pipeline} \label{method-pipleline}
The pipeline of TTS-based simulation can be divided into following steps:
\textbf{1) Dysfluency injection:} for VCTK-Pro, we convert VCTK~\cite{yamagishi2019cstr-vctk} text into IPA sequences via the VITS phonemizer, and for AISHELL3-Pro, we convert Mandarin text into pinyin sequences. We then add different types of dysfluencies at the phoneme/word(English) and pinyin (Chinese) level according to the \textit{TTS rules} (Sec. \ref{tts-rules}).
\textbf{2) VITS~\cite{kim2021conditional-vits} inference: } We take dysfluency-injected IPA/Pinyin sequences as inputs, conduct the VITS inference procedure and obtain the dysfluent speech. 
\textbf{3) Annotation:} We retrieve phoneme alignments from VITS duration model, annotate the type of dysfluency on the dysfluent region. 

\vspace{-5pt}
\subsection{Co-Dysfluency TTS rules} \label{tts-rules}

\begin{figure}[ht]
    \centering
    \includegraphics[height=6.6cm]{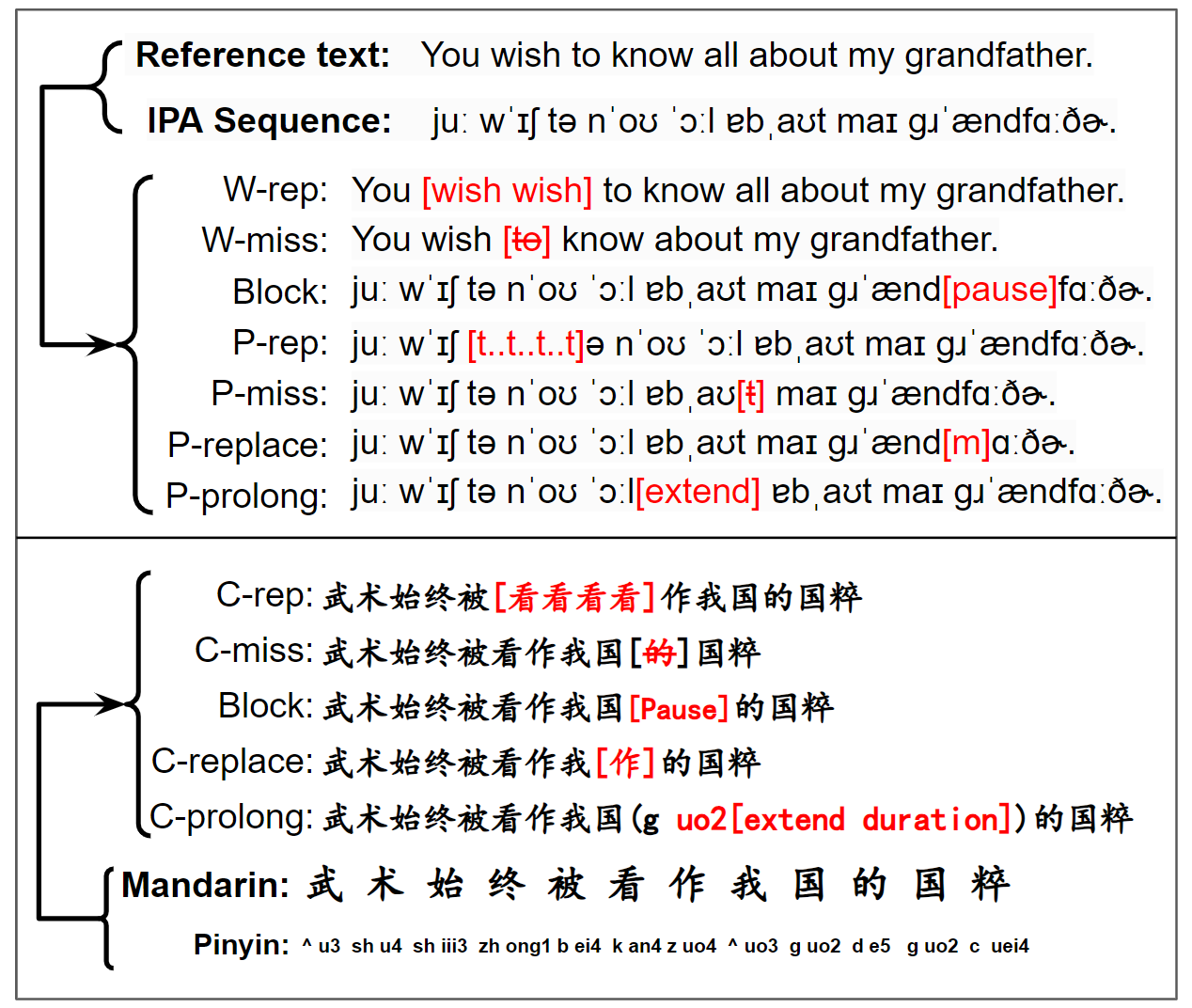}
    \caption{TTS rules for VCTK-Pro and AISHELL3-Pro}
    \label{tts-rules-2lang}
\end{figure}

For VCTK-Pro, we incorporate phoneme and word-level dysfluency; for AISHELL3-Pro, we introduce character-level dysfluency. Dysfluencies are simulated via TTS rules~\cite{zhou2024yolo}, with examples provided in Fig. \ref{tts-rules-2lang}. 

In VCTK-Pro, we introduce \textbf{co-dysfluency}, adding multiple dysfluencies into a single utterance. Co-dysfluency is categorized into single-type and multi-type.
For single-type, we insert 2-3 instances of the same type of dysfluency (involves every type mentioned above) at various positions within an utterance.
For multi-type, we incorporate 5 combinations of dysfluencies: (rep-missing), (rep-block), (missing-block), (replace-block) and (prolong-block), with 2 random positions chosen for each combination within the utterance. Note that due to ethic concerns, de-identification techniques~\cite{gao2021detection} might also be involved in the process.

The statistics of three simulated datasets are detailed in Table. \ref{statistics}.

\vspace{-5pt}
\begin{table}[h]
    \caption{Statistics of simulated datasets (hours)}
    \vspace{-3pt}
    \label{dysfluency-stats-vctk-tts}
    \centering
    \setlength{\tabcolsep}{10pt} 
    \renewcommand{\arraystretch}{1.1} 
    \resizebox{8.3cm}{!}{
    \begin{tabular}{l| c c c c} 
    \toprule
    Dysfluency & \textit{VCTK-Pro} & 
    \textit{VCTK-Art} & \textit{AISHELL3-Pro}  \\
    \hline
    \hline
    Repetition & 258.33 & 111.34 & 102.37\\
    Missing & 180.89  & 107.43  & 100.22\\
    Block & 132.41 & 56.95 & 104.91 \\
    Replace & 128.18 & 56.85 & 100.84 \\
    Prolongation & 87.50  & 59.62 & 103.03\\
    Co-dysfluency & 337.84 & - & - \\
    \midrule
    Total&  1125.1 & 392.19 & 511.37\\
    \bottomrule
    \end{tabular}}
    \label{statistics}
\end{table}
\section{Dysfluency Detection as Object Detection}
\label{sec:detection}
Accurate dysfluency detection necessitates handling text dependencies since stutters are not necessarily monotonic. In this work, we adopt the \textit{soft speech-text alignment} from VITS~\cite{kim2021conditional-vits}, which is one of the SOTA TTS models. Given this speech-text alignment as input, our model requires two main components: an optimal spatial and temporal downsampling method, and an extraction mechanism to accurately attend to the relevant dysfluent signal.
Region-wise dysfluency detection can be viewed as a 1D extension of the 2D object detection problem in computer vision, drawing inspiration from the YOLO~\cite{redmon2016look} method, we design a detector that takes the soft speech-text alignment and produces a fixed size 64 x 8 (temporal dim x output dim) output matrix. At each timestep, 8 values are predicted: dysfluency start \& end bounds, confidence score, and C (=5) class predictions. The detector which utilizes a region-wise prediction scheme consists of spatial pattern collector blocks followed by a temporal analysis unit.
The entire paradigm is shown in Fig. \ref{architecture} and the corresponding modules are detailed in the following.

\vspace{-5pt}
\subsection{Soft speech-text alignments}
\label{sec:vits-alignments}

We obtain $|c_{text}| \times |z|$ monotonic attention matrix $A$ from VITS~\cite{kim2021conditional-vits} that represents how each input phoneme aligns with target speech, where $c_{text}$ is text dimension and $z$ the speech duration. For training, we use the soft alignments $A$ and apply a softmax operation across the text dimension, computing the maximum attention value for each time step. To calculate the soft alignments, we use the original pre-trained text and speech encoders. The former is a Transformer encoder~\cite{vaswani2023attenion} with relative positional embeddings, and the latter is a model that uses non-causal residual blocks used in WaveGlow \cite{Prenger2018WaveglowAF}. This soft-alignment attention matrix is then passed into the detection head for training and inference. 

\subsection{Spatial-Temporal Encoders}
\label{sec:subhead}

We adopt the same spatial-temporal encoders as ~\cite{zhou2024yolo}. It consists of a region-wise spatial encoder and a temporal encoder. 

\subsubsection{Spatial Encoder}
Learnable spatial pattern collector blocks are used to preserve local spatial features. Here, we are going to elaborate more on the intuition. Traditional speech recognition tasks take speech features such as mel spectrograms as input, and the de facto encoder~\cite{gulati2020conformer} is applied. However, the soft alignments $A$ mentioned are spatially different from speech features, such that separate convolutions (pointwise and depthwise) will be ineffective for such input representations. Therefore, a modified convolution paradigm was proposed~\cite{zhou2024yolo}, where a depthwise convolution followed by a grouped convolution, instead of a pointwise convolution, is adopted in this setting. This has been experimentally proven to preserve region-wise information, as visualized in Fig.~\ref{architecture} (b). 

\subsubsection{Temporal Encoder}
Since the task is to predict time-aware dysfluencies, technically it is a region-wise aggregation, which is a 1D sequential timing problem. To achieve this, the transformer encoder~\cite{vaswani2023attenion} is simply applied to handle both global and local timing alignments. We employed transformer-base in this setting.

\subsection{Training Loss}
\label{sec:train-object}
The speech utterance is split into segments of fixed steps. For each segment, we are going to predict three things: (1) the dysfluency confidence score (if and how confident we are that dysfluency exists in this segment), denoted as $y_i$, (2) the boundary of the dysfluencies $b_{\text{start}}$ and $b_{\text{end}}$, and (3) the dysfluency type $c_n$: whether that dysfluency is a block, repetition, replacement, insertion, or missing word.  The bound values are normalized between 0-1 using fixed padded lengths as the max bound values. The balancing factors are $\lambda_{\text{bound}}=5$, $\lambda_{\text{conf}}=1$, and $\lambda_{\text{class}}=0.5$. $S$ is the number of regions and $n$ is the number of classes. $\mathbbm{1}_{obj}$ indicates whether the dysfluency appears in that segment. The loss function is denoted by the following equation:

\parbox{23em}{
\begin{flushright}
$\mathbb{L}=\lambda_{\text{bound}} \frac{1}{S} \displaystyle\sum_{i=0}^{S} \mathbbm{1}_{\text{obj}} [(b_{\text{start}} - \hat{b}_{\text{start}})^2 + (b_{\text{end}} - \hat{b}_{\text{end}})^2]$ \\[-1ex]
$-\lambda_{\text{conf}}  \frac{1}{S} \displaystyle\sum_{i=0}^{S}  \hat{y}_i \log(p(y_i)) + (1 - \hat{y}_i) \cdot \log(1 - p(y_i))$\\[-1ex]
$-\lambda_{\text{class}} \frac{1}{S} \displaystyle\sum_{i=0}^{S} \displaystyle\sum_{j=0}^{n} c_n \log (p(\hat{c}_n))$ 
\end{flushright}
}

\section{Experiments}
\label{sec:experiments}
\subsection{Datasets}
\begin{itemize}[leftmargin=*]

\item \textbf{VCTK~\cite{yamagishi2019cstr-vctk} }includes recordings from 109 native English speakers. Each speaker reads out about 400 sentences from a newspaper, the rainbow passage and an elicitation paragraph used for the speech accent archive. This corpus encompasses about 48 hours of accented speech. It is used for simulating \textit{VCTK-Pro} and \textit{VCTK-Art}.
\item 
\textbf{AISHELL-3~\cite{AISHELL-3_2020}} is a large-scale and high-fidelity multi-speaker Mandarin speech corpus which includes 218 native Chinese mandarin speakers with roughly 85 hours of emotion-neutral recordings. It is used for simualting \textit{AISHELL3-Pro}. 
\item \textbf{LibriStutter~\cite{kourkounakis2021fluentnet}} is a synthesized dataset which contains artificially stuttered speech and stutter classification labels for 5 stutter types. It was generated using 20 hours of audio selected from the `dev-clean-100' section of \cite{panayotov2015librispeech}. 

\item \textbf{UCLASS~\cite{Howell2009TheUA}}
 contains recordings from 128 children and adults who stutter. Only 25 files have been annotated and did not annotate for the block class, we only used those files and did not use the block class for subsequent datasets. 
 
 \item \textbf{SEP-28K} is curated by~\cite{lea2021sep}, contains 28,177 clips extracted from publicly available podcasts. These clips are labeled with five event types including block, prolongation, sound / word repetition and interjection. Clips labeled as ``unsure" in the were excluded from the dataset.
 
 \item \textbf{Aphasia Speech} is collected from our clinical collaborators, our dysfluent data comprises 46 participants (38 English speakers and 8 Chinese mandarin speakers) diagnosed with Primary Progressive Aphasia (PPA), larger than the data used in ~\cite{lian2023unconstrained-udm, lian2024hierarchical} which only has 3 English speakers. 
\end{itemize}

\subsection{Training}
We trained the detector for 30 epochs using a 90/10 train/test split, which was separately applied to three simulated datasets. 
We utilized a batch size of 64 and leveraged the Adam optimizer, configured with beta values of 0.9 and 0.999 and a learning rate of 3e-4. We choose not to use dropout or weight-decay in our training process.
Training on VCTK-Art, VCTK-Pro (without co-dysfluency) and AISHELL3-Pro requires a total of 39, 41 and 36 hours respectively, on a RTX A6000.

\subsection{Metrics}
\begin{itemize}[leftmargin=*]

\item \textbf{Phoneme Error Rate (PER)} is a measure of how many errors (inserted, deleted, and changed phonemes) are predicting phoneme sequences compared to the actual phoneme sequence. It calculated by dividing the number of phoneme errors by the total number of phonemes.

\item  \textbf{Accuracy (Acc.)} refers to the correctness of predictions regarding types of dysfluency within regions that exhibit some form of dysfluency. 

\item \textbf{Bound loss} is calculated as the mean squared error between the predicted and actual boundaries of dysfluent regions within a 1024-length padded spectrogram, which is then converted to a time scale using a 20ms sampling frequency. For co-dysfluency analyses, the bound loss is averaged across all identified dysfluent regions.

\item \textbf{Time F1~\cite{lian2023unconstrained-udm}} measures the accuracy of boundary predictions by assessing the overlap between predicted and actual dysfluent region bounds. A sample is classified as a True Positive if any intersection occurs between these bounds.
\end{itemize}

\subsection{Evaluation of dysfluency simulation}
\subsubsection{MOS tests}
To evaluate the rationality and naturalness of three datasets we constructed, we collected Mean Opinion Score (MOS, 1-5) ratings from 11 people.  The results are displayed in Table. \ref{mos}.
Our three simulated datasets were perceived to be far more natural than the VCTK++~\cite{lian2023unconstrained-udm} baseline corpus. Notably, VCTK-Pro was rated as closely mimicking human speech.

\begin{table}[h]
    \caption{MOS for Simulated datasets}
    \vspace{2pt}
    \label{dysfluency-stats-vctk-tts}
    \centering
    \setlength{\tabcolsep}{10pt} 
    \renewcommand{\arraystretch}{1.2} 
    \resizebox{8.5cm}{!}{
    \small
    \begin{tabular}{l| c c c c} 
    \toprule
    Dysfluency Type &  \textit{VCTK++} & \textit{VCTK-Art} & \textit{VCTK-Pro} & \textit{AISHELL3-Pro} \\
    \hline
    \hline
    Repetition & 1.40 ± 0.70 & 2.61 ± 1.05 & 3.33 ± 0.86 & 3.88 ± 0.73\\
    Missing  & N/A & 3.44 ± 1.23 & 3.89 ± 1.05 & 3.37 ± 1.06\\
    Block  & 2.80 ± 0.63 & 3.35 ± 1.13 & 3.22 ± 1.09 & 2.96 ± 1.03\\
    Replace & N/A & 3.48 ± 1.42 & 2.62 ± 1.21 & 3.13 ± 0.74 \\
    Prolongation & 1.20 ± 0.79& 2.55 ± 0.53 & 3.00 ± 1.00 & 2.64 ± 1.12\\
    \midrule
    Overall & 1.80 ± 0.74 & 3.08 ± 1.12 & \textbf{3.21 ± 0.97} & 3.19 ± 0.95
    \\
    \bottomrule
    \end{tabular}}
    \label{mos}
\end{table}

\vspace{-10pt}
\subsubsection{Dysfluency intelligibility }
In order to verify the intelligibility of simulated datasets, we use phoneme recognition model~\cite{allosaurus} to evaluate the raw VCTK (/) and various types of dysfluent speech from VCTK-Pro and VCTK-Art. Results in Table.~\ref{ctc-result} show generally low PER, indicating good intelligibility and usability despite higher PERs than raw VCTK.
Comparatively, VCTK-Pro performs better overall, while VCTK-Art excels particularly in repetition and block. AISHELL3-Pro was not evaluated due to the lack of available high-quality pinyin-level Chinese speech recognition model.

\begin{table}[h]
    \caption{Phoneme Transcription Evaluation}
    \label{ctc-result}
    \centering
    \setlength{\tabcolsep}{10pt} 
    \renewcommand{\arraystretch}{1.3} 
    \resizebox{8.7cm}{!}{
    \large
    \begin{tabular}{l|c c c c c c} 
     \toprule
      \rowcolor{brightturquoise} 
     \vspace{-2pt}
     & \multicolumn{6}{c}{PER ($\% \downarrow$)} \\
    Type & / & Repetition  & Missing & Block & Replace & Prolongation \\
    \hline
    \hline
    \textit{VCTK-Art} & \multirow{2}{*}{6.243} & \textbf{8.328} & 8.250 & \textbf{10.118} & 9.665 & 9.893\\
    \textit{VCTK-Pro} & & 8.869 & \textbf{7.600} & 11.974 & \textbf{8.004} & \textbf{6.346} \\
    \bottomrule
    \end{tabular}}
\end{table}

\begin{table*}[h]
    \caption{Accuracy (Acc) and Bound loss (BL) of the five dysfluency types trained on the VCTK-Pro and VCTK-Art.}
    \label{dysfluency-eval-train}
    \centering
    \setlength{\tabcolsep}{8pt} 
    \renewcommand{\arraystretch}{1.1} 
    \resizebox{17cm}{!}{
    \small
    \begin{tabular}{l c c|c c| c c| c c |c c |c c} 
     \toprule
     & Trainable & & \multicolumn{2}{c|}{Rep}& \multicolumn{2}{c|}{Block} & \multicolumn{2}{c|}{Miss} & \multicolumn{2}{c|}{Replace} & \multicolumn{2}{c}{Prolong}\\
    Methods& parameters & Dataset& Acc.\% & BL & Acc.\%  & BL & Acc.\%  & BL & Acc.\%  & BL & Acc.\%  & BL \\
    \hline
    \hline
    H-UDM~\cite{lian2024hierarchical} & 92M & VCTK-Art & 84.29 & 29ms &97.59 &24ms& 29.11 & 27ms&-&-&-&-\\ 
    Stutter-Solver(VCTK-Art) & 33M & VCTK-Art
    & \textbf{87.55} & \textbf{21ms} & \textbf{99.64} & \textbf{15ms} & \textbf{91.17} & \textbf{12ms} & \textbf{66.81} & \textbf{15ms} & \textbf{79.16} & \textbf{17ms}
    \\
    
    \midrule
    H-UDM~\cite{lian2024hierarchical} & 92M & VCTK-Pro & 74.66 &68ms & 88.44& 85ms&15.00&100ms &-&-&-&-\\
    Stutter-Solver(VCTK-Pro) & 33M & VCTK-Pro & \textbf{98.78} & \textbf{27ms} & \textbf{98.71} & \textbf{78ms} & \textbf{70.00} &  \textbf{8ms} & \textbf{73.33} & \textbf{10ms} &  \textbf{93.74} & \textbf{12ms}\\
    \midrule
    Stutter-Solver(AISHELL3-Pro) & 33M & AISHELL3-Pro & 93.33 & 17ms & 99.98 & 52 ms & 95.00 & 2ms & 95.16 & 13ms & 96.55 & 16ms\\
    \bottomrule
    \end{tabular}}
\end{table*}

\begin{table}[htp!]
    \caption{Dysfluency evaluation on Aphasia speech.}
    \label{dysfluency-eval-ppa}
    \centering
    \setlength{\tabcolsep}{6pt} 
    \renewcommand{\arraystretch}{1.2} 
    \resizebox{8.7cm}{!}{
    \small
    \begin{tabular}{l c c c} 
    \toprule
    Methods & Ave. Acc. ($\%$, $\uparrow$) & Best Acc.($\%$, $\uparrow$) & Ave. BL (ms, $\downarrow$)\\
    \hline
    \hline
    \vspace{0.5mm}
    H-UDM~\cite{lian2024hierarchical} &41.8 & 70.22&52ms\\
    Stutter-Solver(VCTK-Art) & 52.82 & \textbf{93.47 (Repetition)} & 41ms  \\
    Stutter-Solver(VCTK-Pro)  & 54.19 & \textbf{92.54 (Block)} & 21ms\\
    Stutter-Solver(AISHELL3-Pro) & 72.37 & \textbf{94.85 (Block)} & 13ms\\
    \bottomrule
    \end{tabular}}
\end{table}

\vspace{-5pt}
\subsection{Dysfluency detection}
To assess the performance of trained detector, we conduct evaluations on three simulated datasets, as well as on the PPA data. The results, which include type-specific detection accuracy and bound loss metrics, are detailed in Table.~\ref{dysfluency-eval-train} for the simulated datasets and in Table.~\ref{dysfluency-eval-ppa} for the PPA data. 
Additionally, we compared our results with previous works by validating it on UCLASS, Libristutter, and SEP-28K, where we computed type-specific accuracy and Time F1, as shown in Table.~\ref{compare-with-benchmark}.

In Table.\ref{dysfluency-eval-train}, we used H-UDM~\cite{lian2024hierarchical} as the baseline. Both versions of Stutter-Solver(VCTK-Art and VCTK-Pro) surpassed H-UDM across all metrics. Notably, Stutter-Solver(VCTK-Pro) showed stronger results for English. Additionally, AISHELL3-Pro performed even better, likely due to the unique pronunciation traits of Chinese and its noticeable character-level dysfluency.
In Table. \ref{compare-with-benchmark}, we presented our results using publicly available datasets (UCLASS, LibriStutter, and SEP-28K). Since the original benchmarks use private test sets, direct accuracy comparisons may not be completely fair. We instead emphasized time-aware detection, reporting the Time F1 score for each dataset. All baselines, except H-UDM, scored 0. Our proposed methods consistently outperformed H-UDM in these evaluations.
In Table. \ref{dysfluency-eval-ppa}, both versions of Stutter-Solver outperformed H-UDM, and the Chinese model performed best on Chinese PPA. However, the average accuracy remained low, underscoring the challenge of accurately capturing the real distribution of dysfluency.

\begin{table}[h]
    \caption{Type-specific accuracy (ACC) and time F1-score}
    \label{compare-with-benchmark}
    \centering
    \setlength{\tabcolsep}{5pt} 
    \renewcommand{\arraystretch}{1} 
    \resizebox{8.8cm}{!}{
    \small
    \begin{tabular}{l c| c c c| c} 
     \toprule
    Methods & Dataset & \multicolumn{3}{c|}{Accuracy ($\%$, $\uparrow$)} & Time F1 ($\uparrow$)\\
    \rowcolor{brightturquoise}
    & & \textit{Rep} & \textit{Prolong} & \textit{Block} &\\ 
    \midrule
    Kourkounakis et al. ~\cite{kourkounakis2021fluentnet}& UCLASS & 84.46 & 94.89 & - & 0\\
    Jouaiti et al. ~\cite{segment-detection4} & UCLASS & 89.60 & 99.40 & - & 0\\
    Lian et al. ~\cite{lian2024hierarchical} & UCLASS & 75.18&- & 50.09 & 0.700\\
     \textbf{Stutter-Solver (VCTK-Art)} & UCLASS &82.56 & 84.83 & \textbf{64.42} & 0.806\\
    \textbf{Stutter-Solver (VCTK-Pro)} & UCLASS & \textbf{92.00} & 91.43 & 56.00  & \textbf{0.893}\\
    \midrule
    Kourkounakis et al. ~\cite{kourkounakis2021fluentnet}& LibriStutter & 82.24& 92.14 & - & 0\\
    Lian et al. ~\cite{lian2024hierarchical} &LibriStutter &85.00& -& -& 0.660\\
     \textbf{Stutter-Solver (VCTK-Art)} & LibriStutter &89.04 & 62.58 & - & 0.686\\
    \textbf{Stutter-Solver (VCTK-Pro)} & LibriStutter& \textbf{89.71} & 67.74& - & \textbf{0.697}\\
    \midrule
   Jouaiti et al. ~\cite{segment-detection4} &SEP-28K& 78.70& 93.00 & - & 0 \\
    Lian et al. ~\cite{lian2024hierarchical} &SEP-28K & 70.99& -&66.44&0.699 \\
     \textbf{Stutter-Solver (VCTK-Art)} & SEP-28K & 
     78.31 &  74.99 & 68.02 & 0.786 \\
    \textbf{Stutter-Solver (VCTK-Pro)} & SEP-28K& \textbf{82.01}&  89.19& \textbf{68.09} & \textbf{0.813}\\    
    \bottomrule
    \end{tabular}}
\end{table}

\vspace{-12pt}
\subsection{Co-dysfluency}
In Section \ref{tts-rules}, we incorporated \textbf{co-dysfluency} into VCTK-Pro. We trained \textit{Stutter-Solver} on single-type, multi-type, and mixed-type (single \& multi) co-dysfluency respectively, and measured average accuracy and bound loss using corresponding simulated data.
The results, shown in Table. \ref{table:co-dysfluency}, demonstrate that our detector's performance on co-dysfluency matches its capability in simpler scenarios with only one dysfluency per utterance.
This indicates that our detector handles co-dysfluency effectively. It is worth noting that this fundamental property is missing in previous work. 

\begin{table}[htp!]
    \caption{Evaluation of Co-dysfluency}
    \vspace{-5pt}
    \label{table:co-dysfluency}
    \centering
    \setlength{\tabcolsep}{6pt} 
    \renewcommand{\arraystretch}{1.2} 
    \resizebox{8.8cm}{!}{
    \small
    \begin{tabular}{l c c c} 
    \toprule
    Methods & Co-dysfluency & Ave Acc.($\%$, $\uparrow$) & Ave. BL (ms, $\downarrow$)\\
    \hline
    \hline
    Stutter-Solver(/)  & / & \textbf{91.24} & 29ms\\
    \hline
    Stutter-Solver(Single-type) & Single-type & 90.22 & 26ms\\
    Stutter-Solver(Multi-type) & Multi-type & 89.59 & \textbf{15ms} \\
    Stutter-Solver(Mix-type) &  Mix-type & 90.08 & 24 ms\\
    \bottomrule
    \end{tabular}}
\end{table}

\begin{table}[htp!]
    \caption{Evaluation of Multi-lingual}
    \vspace{-5pt}
    \label{table:multi-lingual}
    \centering
    \setlength{\tabcolsep}{6pt} 
    \renewcommand{\arraystretch}{1.2} 
    \resizebox{8.6cm}{!}{
    \small
    \begin{tabular}{l c c c} 
    \toprule
    Methods & Dataset & Ave Acc.($\%$, $\uparrow$) & Ave. BL (ms, $\downarrow$)\\
    \hline
    \hline
    Stutter-Solver(VCTK-Pro)  & VCTK-Pro & 91.24 & 29ms\\
    Stutter-Solver(Multi-lingual) & VCTK-Pro & \textbf{93.98} & \textbf{21ms} \\
    \hline
    Stutter-Solver(AISHELL3-Pro) & AISHELL3-Pro & \textbf{96.00} & \textbf{20ms}\\
    Stutter-Solver(Multi-lingual) & AISHELL3-Pro & 86.88 & 43ms\\
    \bottomrule
    \end{tabular}}
\end{table}

\subsection{Multi-lingual}
\vspace{-5pt}
In addition to training the detector separately on single languages, we trained it simultaneously on two languages to evaluate its performance in a multi-lingual scenario. 
We randomly sampled 300 hours of data from both VCTK-Pro and AISHELL3-Pro for training. The results, presented in Table. ~\ref{table:multi-lingual}, show that multi-lingual training slightly improved detection performance for English but significantly reduced it for Chinese compared with training separately on a single language. This indicates that multi-lingual training has varying effects on detection accuracy depending on the language. It is important to note that our method does not require additional language-specific dysfluency templates, in contrast to the previous state-of-the-art work by~\cite{lian2024hierarchical}.

\vspace{-5pt}
\section{Conclusions and limitations}
\label{sec:conclusion}
\vspace{-5pt}
We propose \textit{Stutter-Solver} that detects speech dysfluencies in an end-to-end manner. \textit{Stutter-Solver} is able to handle co-dysfluencies within the utterance and is a natural multi-lingual dysfluency detector. We proposed three annotated dysfluency simulated corpora such that \textit{Stutter-Solver} achieves state-of-the-art performance on a couple of dysfluency corpora. However, limitations exist. First, the performance on real nfvPPA speech is far worse than that on simulated speech. Future work will focus on reducing the gap between simulated and real dysfluency distributions. Second, the proposed simulated corpora are not at a large scale and we have not reached the limit. Future work will focus on pushing the limit of scaling efforts when more data and resources are available. Third, it is worth exploring the simulation in gestural space~\cite{lian22bcsnmf, lian2023factor} or rtMRI space~\cite{wu23k_interspeech} instead of articulatory EMA space for finer-grained control. It is also worth exploring both speaker-dependent and speaker-independent dysfluencies via disentangled analysis and synthesis~\cite{lian2022utts, qian2022contentvec, lian2022robust-d-dsvae, lian2022towards-c-dsvae, choi2022nansy++}, which serves as the foundation for behavioral dysfluency study.

\section{Acknowledgement}
Thanks for support from UC Noyce Initiative, Society of Hellman Fellows, NIH/NIDCD and the Schwab Innovation fund.



\newpage
\bibliographystyle{IEEEbib}
\bibliography{refs}

\end{document}